\documentclass[aps,amsmath,twocolumn,amssymb,prl]{revtex4-1}

\usepackage{graphicx}
\usepackage{amsmath}

\usepackage{hyperref}
\usepackage[all]{hypcap}
\hypersetup{
	colorlinks=true,		
	linkcolor=blue,		
	citecolor=red,		
	filecolor=magenta,		
	urlcolor=cyan		
} 

\begin{document}

\author{B. J. Moehlmann}
\author{M. E. Flatt\'e}
\affiliation{Optical Science and Technology Center and Department of Physics and Astronomy, University of Iowa, Iowa City, Iowa, 52242}
\title{Anholonomic spin manipulation in drift transport in semiconductors}

\begin{abstract}
We find that the electronic spin rotation induced by drift transport around a closed path in a wide variety of nonmagnetic semiconductors at zero magnetic field depends solely on the physical path taken.  Physical paths that produce any possible spin rotation due to  transport around a closed path are constructed for electrons  experiencing strain or electric fields in (001), (110), or (111)-grown zincblende semiconductor quantum wells. Spin decoherence due to travel along the path is negligible compared to the background spin decoherence rate.  The small size of the designed paths ($< 100$~nm scale in GaAs) may lead to applications in nanoscale spintronic circuits.
\end{abstract}

\maketitle

Anholonomy, in which a vector transported around a closed curve in parameter space does not return to its initial orientation, is a general property of quantum mechanical systems, and has analogs in other fields including optics and classical mechanics.\cite{Shapere1989}
Within quantum mechanics the geometrical anholonomic phase, which depends only on the path taken and not on the time taken to traverse the path, is the Berry phase.\cite{Berry1984} Recently the spin precession during drift transport of polarized packets of electron spin in semiconductors has been shown to be independent of transit time when no magnetic field is applied.%
\cite{Schliemann2003
	,Kato2004c
	,Crooker2005a
	,Kato2005b
} The source of this behavior is the special nature of the spin-orbit Hamiltonian in these systems, where the lack of inversion symmetry combined with  strain
\cite{Kato2004c
	,Crooker2005a
	,Kato2005b
} or confinement in a quantum well\cite{Schliemann2003} produces a spin splitting that depends linearly on the electron crystal momentum. Spin precession that is independent of transit time, however, is not sufficient to demonstrate anholonomic transport and a geometrical phase, for reversal of the path taken will unwind the spin precession such that the electron spin will return to its original orientation. Anholonomic manipulation of spin in quantum dots has been described\cite{Coish2006,San-Jose2008,Golovach2010}, however this effect requires coherent manipulation of the confined electron's wavefunction and does not apply to spin manipulation of an orbitally-incoherent ensemble of spins undergoing drift transport. 

Here we demonstrate that the drift of electron-spin polarization in a semiconductor, under the influence of strain or quantum well confinement, corresponds to a general nonholonomic system, whereby closed paths can be constructed that will produce any desired spin precession on an initial spin, depending only on the path and not on the transit time. Ordinarily in an anholonomic system the Berry phase must be separated from the dynamical phase, which does depend on the transit time; in these nonmagnetic semiconductor systems the dynamical phase vanishes, leaving only the Berry phase (an SU(2) rotation of the spin) behind. Corrections to the linear dependence of the spin splitting on the electron crystal momentum produce dynamical phases, and also spin decoherence. We identify realistic situations where these corrections are small. Typical scales of the paths in 10~nm quantum wells are $50-2500$~nm, leading to the potential for sub-$100$~nm scale spin manipulation elements. Anholonomic spin phenomena will permit the preparation of arbitrarily-oriented spin packets for probing fundamental spin dynamics as well as new methods of electrically controlling spin resonance.  Current interest in spin manipulation for information processing\cite{Awschalom2002,Awschalom2007} may motivate the use of this effect for spintronic devices.


Spin anholonomy is possible with any of the typical spin-orbit fields generated in semiconductor quantum wells, including the so-called Rashba, Dresselhaus and strain-induced fields, and for (001), (110), and (111) quantum wells.  
Each of these spin-orbit interactions can be described by a precession field ${\ensuremath{	\mathbf \Omega{(\mathbf{k})}}}=-{\ensuremath{	\mathbf \Omega{(-\mathbf{k})}}}$ which depends linearly on the crystal momentum ${\bf k}$, and the Hamiltonian 
\begin{equation}
H_{so} = {\vec{S}} \cdot \mathbf\Omega(\mathbf{k}) = \frac{\hbar}{2} \mathbf\sigma \cdot \mathbf\Omega(\mathbf{k}) = \frac{\hbar}{2} \sigma \cdot \Xi \cdot {\mathbf k} \label{hso}
\end{equation}
for the electrons of the lowest-energy conduction band, where $\Xi$ is a second-rank tensor (the left-side dot product in Eq.~(\ref{hso}) involving one set of indices and the right-side dot product the other). The effect of a Hamiltonian such as Eq.~(\ref{hso}) on spin precession in transport has been observed\cite{Kato2004c,Crooker2005a,Beck2006,Crooker2007} and utilized to initialize and manipulate spins through control of the transport direction\cite{Kato2005}. The correspondence between precession lengths in transport and forms of the Hamiltonian has been extensively investigated\cite{Bernevig2005,Hruska2006,Yang2008}. 

We describe the motion of a population of carriers with uniform total density, but a spatially and temporally-dependent spin polarization using the semiclassical Boltzmann equation\cite{Qi2003}:
\begin{equation}	
\partial_{t} \hat{F} + \frac{ \hbar }{ {m^*} } \mathbf{k} \cdot \mathbf\nabla_{\mathbf r} \hat{F} + \frac{e }{\hbar}\mathbf{E}\cdot\mathbf\nabla_{\mathbf{k}}\hat{F} + \frac{1}{i\hbar}
\left[{H_{so}},{\hat{F}}\right]
  = -\left( \frac{\partial \hat{F}}{\partial t} \right)_{c} . \label{eq:SCBE}
\end{equation}
The condition of uniform total density can be enforced by introducing an Ansatz for the spin-dependent density matrix $\hat{F}$, such that 
$\hat{F}(\mathbf{x},\mathbf{k},t) = ({\rm I} + {\bf m}(\mathbf{x},t)\cdot\mathbf\sigma) f(\mathbf{k})$, 
where the trace of $f(\mathbf{k})$ over  $\mathbf{k}$ yields the uniform density $n_0$,  ${\rm I}$ is the identity matrix, and ${\bf m}(\mathbf{x},t)$ is the local (dimensionless) magnetization.  We then make the relaxation time approximation, introducing ${\rm I} [f({\bf k})-f_0({\bf k})]/\tau$ for the spin-independent portion of the collision term on the right hand side of Eq.~(\ref{eq:SCBE}) (where $f_0(\mathbf{k})$ is the equilibrium orbital occupation function and $\tau$ is the orbital relaxation time) and $({\bf m}(\mathbf{x},t)\cdot\mathbf\sigma) f(\mathbf{k})/T_{1}$ for the (isotropic) spin-dependent portion of the collision term. Anisotropic spin relaxation can be described with anisotropic relaxation times $T_1$ for the different magnetization directions.  When this Ansatz is introduced, and the resulting equation is traced over spin, the ordinary spin-independent steady-state Boltzmann equation is obtained, $(e{\bf E}/\hbar)\cdot \nabla_{\bf k} f({\bf k}) = -[f({\bf k}) - f_0({\bf k})]/\tau$.

By taking the trace of Eq.~(\ref{eq:SCBE}) with different Pauli matrices, and using the definition of the (uniform) charge current ${\bf j}_0 = n_0e\mu {\bf E} =  {\rm Tr}_{\mathbf{k}}(e\hbar \mathbf{k}/m^*)f(\mathbf{k})$, we find
\begin{equation}
\frac{\partial {\bf m}(\mathbf{x},t)}{\partial t}  
= -\mu({\bf E} \cdot \nabla_\mathbf{x}) {\bf m}(\mathbf{x},t) 
- \frac{ \mu m^*}{\hbar}{\bf m}(\mathbf{x},t) \times \Xi\cdot{\bf E} 
- \frac{ {\bf m}(\mathbf{x},t)}{T_1}.\label{spacetime}
\end{equation}
We note two special cases of particular relevance to spin transport. If the spin polarization is uniform in space, then the $\nabla_\mathbf{x} {\bf m}(\mathbf{x},t)$ term on the right hand side vanishes and we find a direct relationship between the electric field and the temporally-uniform precession rate of the spatially-uniform spin polarization. For our second case, the steady-state solution, the left hand side vanishes and in the absence of spin relaxation we obtain
\begin{equation}
(\hat {\bf E}\cdot \nabla_\mathbf{x}) {\bf m}(\mathbf{x},t) + \frac{m^*}{\hbar} {\bf m}(\mathbf{x},t) \times \Xi \cdot \hat {\bf E} =0,\label{anholonomy}
\end{equation}
where $\hat {\bf E} = {\bf E}/E$. Eq.~(\ref{anholonomy}) describes propagation that is independent of time and of the strength of the electric field (only depending on the direction of the field). Furthermore, the gradient of ${\bf m}$ is perpendicular to ${\bf m}$, so (as we have neglected spin relaxation), the magnetization's magnitude does not change with time.

 Eq.~(\ref{anholonomy}) is a central result of this work, for it shows that the evolution of a local magnetization of spin-polarized electrons can be described fully by the evolution along a path of the position of those electrons. If the electric field direction is changed with time, Eq.~(\ref{anholonomy}) will hold except during those transient periods during which the electric field direction is changing or the occupation function $f(\mathbf{k})$ is changing to its new steady-state value. For an abrupt change in ${\bf E}$, the timescale of this change will be the orbital relaxation time $\tau$, so for Eq.~(\ref{anholonomy}) to dominate the spin transport, $\tau$ must be much less than the length of time the electric field is constant as the spins are transported around a path.


The precise form of $\Xi$ originates from a variety of physical configurations of the semiconductor, including strain and quantum-well confinement. We use a notation capable of describing all ($k$-linear) precession fields in (001), (110), and (111) quantum wells, with their accompanying possible as-grown strain: 
\begin{equation} \label{standard}
\hbar \mathbf{\Omega}(\mathbf k) = \alpha ( k_{y} \hat{\mathbf{x}} -k_{x} \hat{\mathbf{y}} ) + \beta (k_{x} \hat{\mathbf{x}} - k_{y} \hat{\mathbf{y}}) + \eta k_{y} \hat{\mathbf{z}}.
\end{equation}	
Here $\alpha$ is the  Rashba coefficient (for (111) wells there are other contributions to $\alpha$). In an unstrained (001) quantum well,  $\beta$ is the linear Dresselhaus coefficient and $\eta=0$, whereas for (110) growth $\eta$ is the linear Dresselhaus coefficient and $\beta=0$, and for (111) growth $\eta=\beta=0$. Strain from growth on a lattice-mismatched subtrate modifies the values of $\alpha$, $\beta$, and $\eta$.
It is convenient to express Eq.~\eqref{standard} in a form that will lend itself to the path description developed below,
\begin{equation}
\hbar\mathbf{\Omega}(\mathbf k) = \kappa k \left[ \hat{\mathbf{x}}\cos{(\rho-\phi)} -\hat{\mathbf{y}}\sin{(\rho+\phi)}  + \eta \hat{\mathbf{z}} \sin\phi \right],
\end{equation}	
 where $\kappa^2 = \alpha^2 + \beta^2$, $\tan \rho = \alpha/\beta$, $\hat{\mathbf{z}}$ is perpendicular to the plane of the quantum well, and $k$ and $\phi$ are the magnitude and azimuthal angle of the electron momentum (confined to the quantum well plane).
$\Xi$ is then
\begin{equation}
\Xi_{ij} = \frac{\kappa}{\hbar}
\begin{pmatrix}
\cos\rho	&	\sin\rho	&	0	\\
-\sin\rho	&	-\cos\rho	&	0	\\
0		&	\eta/\kappa	&	0
\end{pmatrix}.
\end{equation}
For example, when $\alpha = \eta = 0$ in Eq.~(\ref{standard}), $\mathbf\nabla m_x =  ({m^*} \beta/\hbar^2) m_z \hat{\mathbf{y}}$, $\mathbf\nabla m_y =  ({m^*} \beta/\hbar^2) m_z \hat{\mathbf{x}}$, and $\mathbf\nabla m_z =  - ({m^*} \beta/\hbar^2) (m_x \hat{\mathbf{y}} + m_y \hat{\mathbf{x}})$.

In general, motion along a straight path yields a spin rotation determined by the direction of the path, $\phi$, and the distance travelled, $r$.  To simplify the description we consider $\eta=0$  and find from Eq.~(\ref{anholonomy}) the electron spin rotation angle
\begin{equation} \label{eq:chiPhi}
\chi = \frac{m^* \kappa r}{\hbar^2} ( 1+ \sin2\phi\sin2\rho )^{1/2},
\end{equation}
and the (in-plane) azimuthal angle of the precession axis
\begin{equation}
\Phi = -\tan^{-1} \left(\frac{\sin(\rho+\phi)}{\cos(\rho-\phi)} \right).\label{eq:chiPhi2}
\end{equation}	
The same type of result holds for $\eta\ne0$
 and $\beta=0$,
although the precession axis lies in a plane oblique to the growth plane.
The length for the spin to precess $2\pi$ (precession length),
\begin{equation}
\Lambda_{\rho}(\phi) = \frac{2\pi\hbar^2}{m^*\kappa} ( 1+  \sin2\phi\sin2\rho)^{-1/2}. 
\end{equation}
In both the pure Dresselhaus ($\rho = 0$) and the pure Rashba ($\rho=\pi/2$) cases the precession is isotropic, thus $\chi = m^{*} \kappa r / \hbar^2$ and $\Lambda_{0}(\phi) = \Lambda_{\pi/2}(\phi) = 2\pi \hbar^2 /(m^{*} \kappa)$. For other $\rho$ the precession length varies with the direction of travel ($\phi$) in the  quantum well plane. 

The spin rotation when traveling along a path can be constructed by sequentially solving Eq.~(\ref{anholonomy}), if the change in $f(\mathbf{k})$ at the points between line segments of the path does not lead to additional spin precession or spin relaxation. This condition can be met if the orbital relaxation time $\tau$ is much shorter than the path transit time (which is also a condition of the relaxation time approximation). We describe the motion around a path as a product of  spin precession operators $\hat{U}_n\ldots\hat{U}_2\hat{U}_1$, where each
\begin{equation}
\hat{U} = \exp\left[	i m^* \kappa r \sqrt{1+\sin2\phi\sin2\rho} \left( \sigma_{x} \cos\rho + \sigma_{y} \sin\rho \right) / \hbar^2
				\right],	
\end{equation}
arises from  the displacement $r$ with azimuth $\phi$ that corresponds to one segment of the path.

We now describe path constructions to generate an arbitrary spin rotation about any axis (the operator $R({\bf \Omega})$), where the desired precession vector ${\bf \Omega}$ is described by  $\chi$, $\Phi$ and $\Theta$. We focus on closed paths, because it is possible to transport a spin an arbitrary distance without precession by combining line segments made up of whole numbers of precession lengths. We find that a path with at least three non-collinear sides $R({\bf \Omega}) =\hat{U}_3 \hat{U}_2 \hat{U}_1$ can produce a general rotation without net spatial displacement, although the solution is not unique. 

\begin{figure}
\begin{center}
\includegraphics[width=\columnwidth]{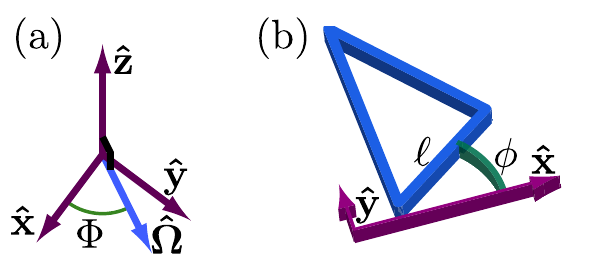}
\caption{\label{fig:inplane axis}	(color online) Sketch of a path which will rotate the spin polarization of an electron about any in-plane axis $\hat{\mathbf\Omega}$ with azimuthal angle $\Phi$. Unlabeled legs are $n\Lambda_\rho$ in length, where $n$ is an integer, $\phi$ is the azimuthal angle of the principal leg.
}
\end{center}
\end{figure}

We begin with a rotation about an in-plane axis. This rotation can be made with a single line segment $\hat{U}_1$, however to achieve spin manipulation with zero net displacement it is necessary return to the origin by a different route, with each leg causing a $2\pi$ rotation. For rotation axes in the quantum well plane ($\Theta=0$), constraining $\hat{U}_3=\hat{U}_2 = \rm{I}$ (by making them $\Lambda$ in length) and minimizing the total length will yield a unique solution, corresponding to  the triangle shown in Fig.~\ref{fig:inplane axis}. 
The principal leg of the triangle is of length $\ell$ and azimuth $\phi$, which are derived from  $\chi$ and $\Phi$ by inverting Eqs.~(\ref{eq:chiPhi}-\ref{eq:chiPhi2}):
\begin{align}
\label{eq:ellphi}
 \ell = &\ \frac{ \hbar^2\chi }{ m^* \kappa \sqrt{ 1+\sin2\rho\sin2\phi}},
 \\ \nonumber
\phi = &\ \cos^{-1}{\left(\frac{	
		\cos{(\rho-\Phi)}			
		}{
		\sqrt{ 1 + \sin2\rho\sin2\Phi }		
		}
	\right)}. 
\end{align}
The remaining vertex of the triangle is placed  at an intersection of the ellipses of constant phase (with radius $\Lambda(\phi')$) centered at each end of the principal leg. The second and third legs of the triangle return the spin packet to the origin having performed only $2\pi$ precessions --- equivalent to identity operations --- about arbitrary axes.

\begin{figure}
\begin{center}
\includegraphics[width=\columnwidth]{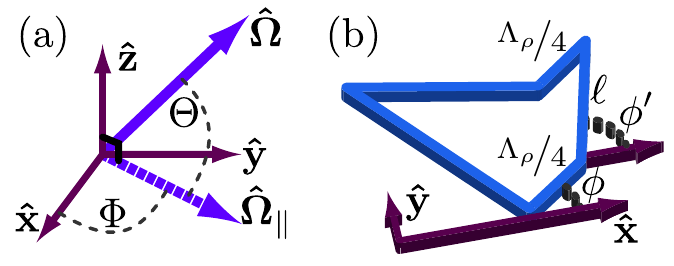}
\caption{	\label{fig:outplane axis}	(color online) Sketch of a path which will rotate the spin polarization of an electron about any arbitrary axis $\hat{\mathbf\Omega}$ with azimuthal angle $\Phi$ and elevation $\Theta$. The projected axis of rotation $\hat{\mathbf\Omega}_\parallel$ with azimuth $\Phi$ is also shown. The auxiliary legs have length $\Lambda_\rho(\phi)/4$ and azimuth $\phi$ from Eq.~(\ref{eq:ellphi}).  The principal leg has length $\ell$  and azimuth $\phi'$ calculated from Eq.~(\ref{eq:ellphi}) using $\Phi' = \Phi-\Theta$ in place of $\Phi$.
 Unlabeled legs are $n\Lambda_\rho$ in length.
}
\end{center}
\end{figure}

Controlled rotation of the spin about an arbitrary (non-planar) axis is also possible. 
	The precession field is always confined to a plane (the quantum well plane in (001) and (111) wells, but an oblique
 plane for (110) wells with a Rashba field), however successive non-collinear displacements result in non-commuting rotations of the spin, which can generate out-of-plane and arbitrary direction rotations. As a special case, for coherent transport (not drift transport as considered here), a square loop in (001) quantum wells causes rotations proportional to the side length about the (001) axis\cite{Hatano2007,Yang2008,Chen2008}.  
An arbitrary rotation axis can, in principal, be implemented by a triangular path, but it is more intuitively implemented by the five sided path shown in Fig.~\ref{fig:outplane axis}. The principal segment, of length $\ell$, is bracketed by two legs of length $\Lambda_\rho/4$; the first rotates the precession axis $\hat{\mathbf{\Omega}}$ into the plane of possible rotation axes and the second restores it to its original orientation. The remaining legs of the path then return the packet to the origin with identity operations on the spin. 

Use of such paths in devices offer great flexibility.  For example, by adjusting the Rashba field with an applied electric field so that $\rho=\pi/4$ (corresponding to the spin helix condition\cite{Bernevig2006}) any motion of a spin around a closed path yields no spin manipulation. Changing the Rashba field away from $\rho=\pi/4$ provides spin manipulation --- permitting gated on-off control of the spin manipulation.


We continue with specific descriptions of the quantities $\alpha$, $\beta$ and $\eta$ for quantum wells and strained semiconductors, and estimates of the size of these closed paths for several materials and quantum wells (tabulated in Table \ref{tab:precession lengths}).  The area enclosed by the path will typically be near to or less than the square of the precession length, which is inversely proportional to the strength of the precession fields.
There are two strain-induced precession fields, one dependent on the shear strain and the other on the biaxial. They are determined by 
\begin{equation}
\hbar\Omega_x =  C_3 ( k_{y} e_{xy} - k_{z}e_{xz} ) + C_4 k_{x}(e_{yy}-e_{zz})
\end{equation}
and the other components are obtained by cyclic permutation. We now describe the spin precession fields for a quantum well strained by growth on a lattice-mismatched substrate.
In both the (001) and (110) quantum wells $\alpha$ is the Rashba coefficient. For (001) growth $\eta=0$ and
\begin{equation}
\beta = - \gamma \left\langle k_z^2 \right\rangle + C_4 \frac{c_{xxxx}+2c_{xxyy}}{c_{xxxx}} \frac{\Delta a}{a}
\end{equation}
where $c_{ijkl}$ and $a$ are the stiffness tensor and lattice constant for the well material.
For  (110) growth $\beta = 0$ and 
\begin{equation}
\eta = \gamma \left\langle k_z^2 \right\rangle / 2 + (C_3 - C_4) \frac{c_{xxxx}+2c_{xxyy}  }{c_{xxxx}+c_{xxyy} + c_{xyxy}} \frac{\Delta a}{a} .
\end{equation}
For (111) growth $\beta = \eta = 0$ and 
\begin{equation}
\alpha = \alpha_R + \frac{2}{\sqrt 3} \gamma \left\langle k_z^2 \right\rangle + \sqrt3 C_3 \frac{ c_{xxxx}+2c_{xxyy}}{c_{xxxx}+2c_{xxyy}+2c_{xyxy} } \frac{\Delta a}{a} ,
\end{equation}
where $\alpha_R$ is the Rashba coefficient.

\begin{table}[h!]
\begin{tabular}{|rc|cccccccc|}
\hline
$w$	&	$\hat{\mathbf{z}}$	&	GaAs	&	InAs	&	GaSb	&	InSb	&	ZnSe	&	CdSe	&	GaP	&	InP \\\hline
5 nm	&	(001)	&	66  	&	37 		&	1.1 	&	1.2  	&	390		&	660		&	350		&	24 \\
10 nm	&	(001)	&	260  	&	150  	&	43  	&	47  	&	1600 	&	2700 	&	1400 	&	970 \\
10 nm	&	(110)	&	530  	&	290  	&	86  	&	93  	&	3100 	&	5300 	&	2800 	&	1900 \\
10 nm	&	(111)	&	230  	&	130  	&	37  	&	40  	&	1300 	&	2300 	&	1200 	&	840 \\ \hline
\end{tabular}
\caption{ \label{tab:precession lengths}
The precession lengths (in nm) of a series of quantum wells of differing thickness and growth direction composed of various III-V and II-VI semiconductors considering only the Dresselhaus field. $w$ is the depth of the quantum well, and $\hat{\mathbf{z}}$ is the growth orientation for it.
}
\end{table}

Our concluding concern is the role of spin relaxation on the spin transport phenomena described here. Spin relaxation in semiconductors occurs via differential precession of spins in the spin-orbit field (dephasing), followed by spin-independent scattering that produces decoherence\cite{Awschalom2002}, and the relaxation rate is proportional to the electron mobility. In the drift regime the drift momentum of the Fermi sea is much smaller than the typical momentum of relevant states (Fermi momentum in the degenerate state, thermal momentum in the non-degenerate state), so the relaxation rate is not affected by transport of carriers (unless the transport leads to an elevated temperature, for which case the relaxation time at the elevated carrier temperature is the relevant quantity\cite{Sanada2002}). The transport time around the path does depend on the mobility, but also on the electric field, so it should be possible with sufficiently large electric fields to move spins around the path with negligible spin decoherence. For example, the spin relaxation rate for a clean ($\mu\sim 5\times 10^3$~cm$^2$/Vs), 5~nm thick GaAs quantum well at room temperature is 20~ps\cite{Terauchi1999}, which in an electric field of $10^4$~V/cm corresponds to a 10~$\mu$m spin transport length, more than 50 times the path length required to perform a spin operation. The error rate in the spin polarization after traversing this path would thus be $<2$\%.



	We have proposed a set of structures that allow for arbitrary manipulation of the spin orientation of an electron packet moving via drift transport, without magnetic fields or net displacements, in a variety of quantum well structures and geometries. The overall size of the paths the spin packets must traverse can be quite small, although it remains an open question whether this form of spin anholonomy will persist in the ballistic transport regime.

We acknowledge helpful conversations with D. Berman and C. E. Pryor. This work was supported by an ONR MURI.
\bibliographystyle{apsrev}
\bibliography{central-bibliography}
\end{document}